\documentclass{ws-ijmpd}

\begin{document}
\markboth{S. Capozziello et al.}{Strings fluid and the dark side of the universe}

\catchline{}{}{}{}{}

\title{A FLUID OF STRINGS AS A VIABLE CANDIDATE TO THE DARK SIDE OF THE UNIVERSE}

\author{S. Capozziello$^{1}$, V.F. Cardone$^{2}$, G. Lambiase$^{2}$ and A. Troisi$^{1}$}

\address{$^1$Dipartimento di Scienze Fisiche, Universit\`a di Napoli, and INFN, Sezione di Napoli, Compl. Univ. di Monte S. Angelo, Edificio G, Via Cinthia, 80121 - Napoli, Italy}

\address{$^2$Dipartimento di Fisica ``E.R. Caianiello'', Universit\`a di Salerno, and INFN, Sezione di Napoli, Gruppo Collegato di Salerno, via S. Allende, 84081 - Baronissi (Salerno), Italy}

\maketitle

\begin{abstract}

We investigate the possibility that part of the dark matter is not
made out of the usual cold dark matter (CDM) dustlike particles,
but is under the form of a fluid of strings with barotropic factor
$w_s = -1/3$ of cosmic origin. To this aim, we split the dark
matter density parameter in two terms and investigate the dynamics
of a spatially flat universe filled with baryons, CDM, fluid of
strings and dark energy, modeling this latter as a cosmological
constant or a negative pressure fluid with a constant equation of
state $w < 0$. To test the viability of the models and to
constrain their parameters, we use the Type Ia Supernovae Hubble
diagram and the data on the gas mass fraction in galaxy clusters.
We also discuss the weak field limit of a model comprising a
significant fraction of dark matter in the form of a fluid of
strings and show that this mechanism makes it possible to reduce
the need for the elusive and up to now undetected CDM. We finally
find that a model comprising both a cosmological constant and a
fluid of strings fits very well the data and eliminates the need
of phantom dark energy thus representing a viable candidate to
alleviate some of the problems plaguing the dark side of the
universe.
\end{abstract}


\section{Introduction}

Soon after the discovery of cosmic acceleration from the Hubble diagram of the high redshift Type Ia Supernovae (SNeIa) \cite{SNeIa,Riess04}, a strong debate arose in the scientific community about the origin of this unexpected result. An impressive flow of theoretical proposals have appeared, while the observational results were constantly providing more and more evidences substantiating the emergence of a new cosmological scenario. The anisotropy spectrum of the cosmic microwave background radiation (CMBR) \cite{CMBR,WMAP,VSA}, the matter power spectrum determined from the clustering properties of the large scale distribution of galaxies \cite{LSS} and the data on the Ly$\alpha$ emitting regions \cite{Lyalpha} all provide indications that the universe have to be described as a spatially flat manifold where matter and its fluctuations are isotropically distributed and represent only about $30\%$ of the overall content. In order to fill the gap and drive the acceleration, a dominant contribute from a homogeneously distributed negative pressure fluid has been invoked. Usually referred to as {\it dark energy}, the nature and the nurture of this mysteryous component represent a new and fascinating conundrum for theoreticians.

While the models proposed to explain this puzzle increase day by day, the most simple answer is still the old Einstein cosmological constant $\Lambda$. Although being the best fit to a wide set of different astrophysical observations \cite{Teg03,Sel04}, it is nevertheless plagued by two evident shortcomings, namely the {\it cosmological constant problem} and the {\it coincidence problem}. A possible way to overcome these problems invokes replacing $\Lambda$ with a scalar field (dubbed {\it quintessence}) evolving down a suitably chosen self interaction potential \cite{QuintFirst}. Although solving the cosmological constant problem, quintessence does not eliminate the coincidence one since too severe constraints on the potential seem to be needed thus leading to the {\it fine tuning problem} \cite{QuintRev}.

The ignorance of the fundamental physical properties of both dark energy and dark matter has motivated a completely different approach to the problem of cosmic acceleration relying on modification of the matter equation of state (EoS). Referred to as {\it unified dark energy} (UDE) models, these proposals resort to a single fluid with exotic EoS as the only candidate to both dark matter and dark energy thus automatically solving the coincidence problem. The EoS is then tuned such that the fluid behaves as dark matter at high energy density and quintessence (or $\Lambda$) at the low energy limit. Interesting examples are the Chaplygin gas \cite{Chaplygin}, the tachyonic field \cite{tachyon} and the Hobbit model \cite{Hobbit}.

It is worth noting that observations only tell us that the universe is accelerating, but they are not direct evidences for new fluids or modifications of the usual matter properties. It is indeed possible to consider cosmic acceleration as the first signal of the breakdown of the laws of physics as we know them. As a consequence, one has to to give off the standard Friedman equations in favour of a generalized version of them arising from some more fundamental theory. Interesting examples of this kind are the Cardassian expansion \cite{Cardassian} and the Dvali\,-\,Gabadadze\,-\,Porrati (DGP) gravity \cite{DGP} both related to higher dimensional braneworld theories. In the same framework, one should also give off the Einsteinian general relativity and turn to fourth order theories of gravitation replacing the Ricci scalar curvature $R$ in the gravity Lagrangian with a generic function $f(R)$ that have been formulated both in the metric \cite{capozcurv,MetricRn,cct} and Palatini approach \cite{PalRn,lnR,ABF04,ACCF} providing a good fit to the data in both cases \cite{noiijmpd,CCF04}. Actually, it is worth noting that it has been recently demonstrated that, under quite general conditions, it is possible to find a $f(R)$ theory that predicts the same dynamics of a given quintessence model.

Although resorting to modified gravity theories is an interesting and fascinating approach, it is worth exploring other possibilities in the framework of standard general relativity. Indeed, all the approaches we have described are mainly interested in solving the dark energy puzzle, while little is said about the dark matter problem. It is worth remembering that dark matter is usually invoked because of the need of a source of gravitational potential other than the visible matter. Considered from this point of view, it is worth wondering whether dark matter could be replaced by a different mechanism that is able to give the same global effect. Moreover, such a mechanism must not alter the delicate balance between dark matter and dark energy that is needed to explain observations. Indeed, if we abruptly reduce the dark matter content of the universe without altering neither the dark energy term nor the background fundamental properties, we are not able to fit the available astrophysical data. Therefore, it is mandatory to test any proposed mechanism both at galactic and cosmological scales.

In a series of interesting papers \cite{letelier}, Letelier investigated the consequences of changing the properties of the right hand side of the Einstein equations adopting a fluid of strings as source term rather than the usual dust matter. Since such strings are not observed at the present time, it seems meaningful to extend the concept of dust clouds and perfect fluid referred to point particles to the case of strings. In particular, Letelier was able to find exact solutions for the case of a spherically symmetric fluid of strings. It is worth noting that such strings could be of cosmological origin \cite{Vil84} and have thus to be included in the energy budget when investigating the dynamics of the universe. It is important to stress, however, that the strings we are referring to have {\it finite lenght} so that the results obtained for a network of cosmic strings of infinite length cannot be extended to the strings considered by Letelier. In this sense, the fluid of finite length strings we are considering represents a generalization of the dust matter. While in this latter case, the matter particles are considered as pointlike, in the case of a fluid of strings\footnote{Hereafter, by {\it fluid of strings} we mean a {\it fluid of finite length strings}.} the elementary constituents are one dimensional objects with finite length.
A fluid of strings has a profound impact at galactic scales. Indeed, assuming that the string transverse pressure was proportional to its energy density, Soleng \cite{soleng} has demonstrated that the force law is altered thus offering the possibility of solving the problem of the flatness of spiral galaxy rotation curves \cite{rc} in a way similar to the MOND proposal \cite{mond}.

Motivated by these considerations, we explore here the possibility that a part (if not all) of the dark matter may be replaced by a fluid of strings whose effective gravitational action may be considered as the source of the gravitational potential needed to flatten the rotation curves. To this aim, we consider cosmic strings as components of such fluid so that its e.o.s. may be simply parametrized by a constant barotropic factor $w_s = -1/3$. Before discussing the impact at galaxy scales, it is preliminarily needed to investigate the effects at cosmological scales. We thus fit different cosmological models, both with and without such a component, to the available astrophysical data in order to test the viability of our proposal and explore if and how the constraints on the model parameters are affected by the presence of a fluid of strings.

The paper is organized as follows. The models we discuss are described in Sect.\,2, while the matching with observations is presented in Sect.\,3 where we also compare the different models in terms of the information criteria parameters. Sect.\,4 is devoted to the weak energy limit of models comprising standard matter embedded in a fluid of strings and show how the corresponding modified gravitational potential could help in reducing the need for CDM. A summary of the results and of their implications are presented in the concluding Sect.\,5

\section{The models}

The key quantity entering most of the usual astrophysical tests is the Hubble parameter $H$ as a function of the redshift $z$. The position of the first peak of the CMBR anisotropy spectrum as measured by WMAP and balloon\,-\,borne experiments \cite{CMBR,WMAP,VSA} is a strong evidence of a spatially flat universe. Assuming therefore $k = 0$, the Friedman equation for the expansion rate $H = \dot{a}/a$ (with $a$ the scale factor normalized to 1 at today) reads\,:

\begin{equation}
H^2 = \frac{8 \pi G}{3} \rho_T = \frac{8 \pi G}{3} \sum_{i = 1}^{N}{\rho_i}
\label{eq: fried1}
\end{equation}
where $\rho_i$ is the energy density of the $i$\,-\,th fluid and the sum is over the $N$ cosmological fluids which make up the cosmic energy budget. If the fluids are not interacting, a conservation equation for each of them hold\,:

\begin{equation}
\dot{\rho}_i + 3 H (1 + w_i) \rho_i = 0
\label{eq: cons}
\end{equation}
with the dot denoting derivative with respect to the cosmic time $t$ and $w_i = p_i/\rho_i$ the barotropic facto of the $i$\,-\,th fluid. Assuming a cosntant $w_i$, Eq.(\ref{eq: cons}) is easily integrated giving\,:

\begin{equation}
\rho_i(z) = \Omega_i \rho_{crit} (1 + z)^{3 (1 + w_i)}
\label{eq: rhovsz}
\end{equation}
with $\Omega_i \equiv \rho_i(z = 0)/\rho_{crit}$ the present day density parameter of the $i$\,-\,th fluid and $\rho_{crit} \equiv 3H_0^2/8 \pi G$ the present day critical density of the universe and, henceforth, we denote with a subscript $0$ all the quantities evaluated today. Inserting Eq.(\ref{eq: rhovsz}) into Eq.(\ref{eq: fried1}), we get\,:

\begin{equation}
H(z) = H_0 \sqrt{\sum_{i = 1}^{N}{\Omega_i (1 + z)^{3 (1 + w_i)}}} \ .
\label{eq: hubbletot}
\end{equation}
To fully assign the model, we have now to specify what are the ingredients of the cosmic pie and the values of their barotropic factors. According to the standard scenario, there are at least three components contributing to the energy budget, namely baryons, dark matter and dark energy. For the former two fluids, it is $p = 0$, while the dark energy is modelled as a negative pressure fluid with constant equation of state $w < 0$ (as in quiessence models) with $w = -1$ giving the usual $\Lambda$ term. Motivated by the considerations discussed in the introduction, we add a fourth component\footnote{Note that we have not included radiation in the energy budget since its density parameter $\Omega_{rad} \sim 10^{-5}$ makes its contribute today indeed negligible.} to our cosmological models. In order to see whether it is possible to reduce the dark matter content of the universe, we replace a fraction $\varepsilon$ of its energy contribute with a fluid of strings characterized by an equation of state \cite{Vil85}\,:

\begin{equation}
w_s = -1/3
\label{eq: ws}
\end{equation}
so that the energy density of the fluid of strings reads\,:

\begin{equation}
\rho_s(z) = \Omega_s \rho_{crit} (1 + z)^2 = \varepsilon \Omega_{DM} (1 + z)^2
\label{eq: rhos}
\end{equation}
where hereafter the subscripts $(b, DM, s, Q)$ denote quantities referred
to baryons, dark matter, fluid of strings and dark energy respectively. The
dimensionless Hubble parameter $E(z) = H(z)/H_0$ thus finally writes\,:

\begin{eqnarray}
E^2(z) & = & \Omega_b (1 + z)^3 + (1 - \varepsilon) \Omega_{DM} (1 + z)^3 + \nonumber \\
~ & + & \varepsilon \Omega_{DM} (1 + z)^2 + \Omega_Q (1 + z)^{3 (1 + w_Q)}
\label{eq: ez}
\end{eqnarray}
where, because of the flatness condition, it is\,:

\begin{equation}
\Omega_Q = 1 - \Omega_b - \Omega_{DM} \ .
\label{eq: oq}
\end{equation}
Before investigating the consequences of introducing the fluids of strings, it is worth spending some more words on the philosophy underlying our model parametrization. Assuming that the dark matter is (mainly) made out of cold dark matter particles and denoting with the subscript $CDM$ the related quantities, in Eq.(\ref{eq: ez}), we have implicitly made the positions\,:

\begin{equation}
\left \{
\begin{array}{lll}
\Omega_{DM} & = & \Omega_{CDM} + \Omega_s \nonumber \\
~ & ~ & ~ \nonumber \\
\ \ \ \ \ \varepsilon & \equiv & \Omega_s/\Omega_{DM} \nonumber
\end{array}
\right .
\label{eq: defeps}
\end{equation}
so that the parameter $\varepsilon$ gives an immediate feeling of what percentage of CDM may be given away without changing dramatically the dynamics of the universe, i.e. still obtaining a good fit to the available astrophysical data. In a sense, we are trying to reduce the need for dark matter replacing its contribute to the dynamics of the universe with a different kind of fluid having a different barotropic factor. Given our ignorance on the fundamental dark matter properties, there is no a priori reason against changing its equation of state. Moreover, it is also conceivable that the total dark matter turns to be made out of both CDM and the fluid of strings. As a consequence, the total matter density parameter is\,:

\begin{equation}
\Omega_M = \Omega_b + \Omega_{CDM} + \Omega_s = \Omega_b + \Omega_{DM} \ .
\label{eq: om}
\end{equation}
Eq.(\ref{eq: ez}) refers to a cosmological model whose energy density is determined by four different components, namely baryons, dark matter, a fluid of strings and dark energy with constant\footnote{Hereafter, we drop the subscript "Q" from $w$ since this only refers to dark energy, while it has been fixed for both matter (baryons and CDM) and fluid of strings.} $w$. Starting from this general case, we define four different models setting some of the parameters as follows\,:

\begin{itemize}

\item{{\it $\Lambda$CDM}\,: $w = -1, \varepsilon = 0$. This is the popular concordance model  that we consider as a testbed of our likelihood analysis.}

\item{{\it $\Lambda$SDM}\,: $w = -1, 0 \le \varepsilon \le 1$. Here, we still retain the cosmological constant as source of cosmic acceleration, but replace a fraction $\varepsilon$ of the dark matter term with the fluid of strings.}

\item{{\it QCDM}\,: $w \le -1/3, \varepsilon = 0$. Also referred to as {\it quiessence}, this model represents the easiest generalization of the successful $\Lambda$CDM scenario. Note that we do not impose a priori $w > -1$ in order to explore the phantom models that seems to be favoured by the recent SNeIa data \cite{Riess04}.}

\item{{\it QSDM}\,: $w \le -1/3, 0 \le \varepsilon \le 1$. This is similar to the QCDM considered above, but now we allow a fraction $\varepsilon$ of the dark matter to be replaced by the string fluid. A caveat is in order here. The fitting procedure does not choose a priori to decrease $\Omega_{CDM}$ while holding fixed $\Omega_Q$. As such, because of Eq.(\ref{eq: oq}), it is possible that the search for the best fit ends in a region of the parameter space where $\Omega_Q$ rather than $\Omega_{CDM}$ is reduced.}

\end{itemize}
As a general remark, we would like to note that although the four models above may formally be considered as a single one (since the former three are clearly particular cases of the latter one), they significantly differ in their underlying physics. As such, choosing among them is not only a matter of finding which one is in better agreement with the observations, but it is rather a sort of compromise between the capability of fitting the data and the physical justification of the model itself.

\section{Matching with observations}

Comparing model predictions with astrophysical observations is a mandatory test of the viability of the given model and also represents an efficient tool to constraint the characterizing parameters. We first describe the method we employ and then discuss the results of the fitting procedure.

\subsection{The method}

We fit the models described in the previous section to the SNeIa Hubble diagram and the data on the gas mass fraction in galaxy clusters. To take into account both datasets, we maximize the following likelihood function\,:

\begin{equation}
{\cal{L}}({\bf p}) \propto \exp{\left [ - \frac{\chi^2({\bf p})}{2} \right ]}
\label{eq: deflike}
\end{equation}
with\,:

\begin{equation}
\chi^2 = \chi_{SNeIa}^2 + \chi_{gas}^2 + \left ( \frac{t_0 - t_0^{obs}}{\sigma_t} \right )^2 \ .
\label{eq: defchi}
\end{equation}
Three terms enter the above $\chi^2$ definition. The first one refers to the SNeIa Hubble diagram and is given by\,:

\begin{equation}
\chi_{SNeIa}^2 = \sum_{i = 1}^{N_{SNeIa}}{\left [ \frac{\mu(z_i, {\bf p}) - \mu_{obs}(z_i)}{\sigma_i} \right ]^2} \ ,
\label{eq: chisneia}
\end{equation}
where $\sigma_i$ is the error on the observed distance modulus $\mu_{obs}(z_i)$ and the sum is over the number $N_{SNeIa}$ of SNeIa observed. On the other hand, the theoretical distance modulus depends on the set of model parameters {\bf p} and may be computed as\,:

\begin{equation}
\mu(z) = 5 \log{D_L(z)} + 25 \label{eq: distmod}
\end{equation}
with $D_L(z)$ the luminosity distance (in Mpc) given by\,:

\begin{equation}
D_L(z) = \frac{c}{H_0} (1 + z)
\int_{0}^{z}{\frac{d\zeta}{E(\zeta; {\bf p})}} \ .
\label{eq: dl}
\end{equation}
The second term in Eq.(\ref{eq: defchi}) is defined as \cite{fgasbib,fgasapp}\,:

\begin{equation}
\chi_{gas}^2 = \sum_{i = 1}^{N_{gas}}{\left [ \frac{f_{gas}(z_i, {\bf p}) - f_{gas}^{obs}(z_i)}{\sigma_{gi}} \right ]^2}
\label{eq: chigas}
\end{equation}
with $f_{gas}^{obs}(z_i)$ the measured gas mass fraction in a galaxy cluster at redshift $z_i$ and $\sigma_{gi}$ the error. For a given cosmological model, $f_{gas}(z, {\bf p})$ is given by \cite{fgasbib,fgasapp}\,:

\begin{equation}
f_{gas}(z) = \frac{b \Omega_b}{(1 + 0.19 \sqrt{h}) \Omega_M} \left [ \frac{D_A^{SCDM}(z)}{D_A^{mod}(z)} \right ]^{1.5}
\label{eq: fgas}
\end{equation}
where $D_A^{SCDM}$ and $D_A^{mod}$ is the angular diameter distance for the SCDM and the model to be tested respectively. $D_A(z)$ may be evaluated as $D_A(z) = (1 + z)^{-2} D_L(z)$. The constant $b$ in Eq.(\ref{eq: fgas}) takes into account the gas lost because of different astrophysical processes. Following \cite{fgasdata}, that have extensively analyzed the set of simulations in Ref.\,\cite{ENF98}, we set $b = 0.824$.

Finally, in Eq.(\ref{eq: defchi}), we have also included a prior on the age of the universe $t_0$ that may be straightforwardly evaluated for a given set of model parameters as\,:

\begin{equation}
t_0 = t_H \int_{0}^{\infty}{\frac{dz}{(1 + z) E(z; {\bf p})}}
\label{eq: tz}
\end{equation}
where $t_H = 1/H_0 \simeq 9.78 h^{-1} \ {\rm Gyr}$ is the Hubble time.

With the definition (\ref{eq: deflike}) of the likelihood function, the best fit model parameters are those that maximize ${\cal{L}}({\bf p})$. However, to constrain a given parameter $p_i$, one resorts to the marginalized likelihood functions normalized at unity at maximum. The $1 \sigma$ confidence regions are determined by $\Delta \chi^2 = \chi^2 - \chi_0^2 = 1$, while the condition $\Delta \chi^2  = 4$ delimited the $2 \sigma$ confidence regions. Here, $\chi_0^2$ is the value of the $\chi^2$ for the best fit model. Projections of the likelihood function allow to show eventual correlations among the model parameters. In these two dimensional plots, the $1 \sigma$ and $2 \sigma$ regions are formally defined by $\Delta \chi^2 = 2.30$ and $6.17$ respectively so that these contours are not necessarily equivalent to the same confidence level for the single parameter estimates.

In order to reduce the space of parameters to explore, we fix the Hubble constant $h$ (in units of $100 \ {\rm km \ s^{-1} \ Mpc^{-1}}$) to the value determined by Daly \& Djorgovski \cite{DD04} fitting the linear Hubble law to a large set of low ($z < 0.01$) redshift SNeIa\,:

\begin{displaymath}
h = 0.664 {\pm} 0.008 \ .
\end{displaymath}
This value is in good agreement with $h = 0.72 {\pm} 0.08$ reported by the HST key project \cite{HSTKey} based on a combined analysis of several local distance ladder methods. Since we are not interested in constraining $H_0$, in the following analysis, we will set $h = 0.664$ neglecting the small uncertainty. Moreover, to a large extent, the effect of changing $h$ on the results may be easily guessed and does not affect significantly the main results.

The baryon density parameter $\Omega_b$ is constrained by theoretical models of nucleosynthesis and by the observed abundance of light elements. Based on these considerations, Kirkman et al. \cite{Kirk} have estimated\,:

\begin{displaymath}
\Omega_b h^2 = 0.0214 {\pm} 0.0020 \ .
\end{displaymath}
Combining this estimate with the value set above for the dimensionless Hubble constant $h$ and neglecting the small error, we therefore fix $\Omega_b = 0.049$.

\subsection{Results}

We have applied the likelihood procedure described above using the SNeIa Gold dataset \cite{Riess04} and the catalog of relaxed galaxy clusters compiled in Ref.\,\cite{fgasdata}. Moreover, we choose $(t_0^{obs}, \sigma_t) = (13.1, 2.9)$ Gyr as obtained from globular clusters \cite{Krauss} and in agreement with estimates from nucleochronology \cite{Cayrel}.

\begin{table}[ph]
\tbl{Summary of the results of the likelihood analysis of the models discussed in the text. The maximum likelihood value ($bf$) of each quantity is reported, while the 68$\%$ (95$\%$) range is $(bf - \delta_-, bf + \delta_+)$ with $\delta_-$ and $\delta_+$ the first (second) number reported as subscript and superscript respectively. The symbol $(\ast)$ means that the parameter is held fixed. Note that, for the QSDM model, we may give only upper limits on $\varepsilon$.}
{\begin{tabular}{|c|c|c|c|c|c|}
\hline
Id & $\Omega_{DM}$ & $\varepsilon$ & $w$ & $t_0$ (Gyr) & $z_T$  \\
\hline \hline
$~$ & $~$ & $~$ & $~$ & $~$ & $~$ \\

$\Lambda$CDM & $0.270_{-0.002 \ -0.006}^{+0.011 \ +0.016}$ & $0 \ (\ast)$ & $-1 (\ast)$
& $13.27_{-0.11 \ -0.16}^{+0.02 \ +0.06}$ & $0.62_{-0.03 \ -0.04}^{+0.01 \ +0.02}$ \\

$~$ & $~$ & $~$ & $~$ & $~$ & $~$ \\

$\Lambda$SDM & $0.270_{-0.004 \ -0.009}^{+0.007 \ +0.012}$ & $0.59_{-0.15 \ -0.35}^{+0.15 \ +0.30}$ & $-1 \ (\ast)$
& $15.51_{-0.55 \ -1.09}^{+0.59 \ +1.40}$ & $0.96_{-0.11 \ -0.24}^{+0.19 \ +0.26}$ \\

$~$ & $~$ & $~$ & $~$ & $~$ & $~$ \\

QCDM & $0.270_{-0.011 \ -0.016}^{+0.003 \ +0.008}$ & $0 \ (\ast)$ & $-1.28_{-0.08 \ -0.15}^{+0.07 \ +0.14}$
& $14.50_{-0.12 \ -0.26}^{+0.15 \ +0.29}$ & $0.60_{-0.01 \ -0.03}^{+0.02 \ +0.03}$ \\

$~$ & $~$ & $~$ & $~$ & $~$ & $~$ \\

QSDM & $0.270_{-0.007 \ -0.015}^{+0.005 \ +0.009}$ & $0 \ (\le 0.16 \le 0.53)$ & $-1.21_{-0.08 \ -0.17}^{+0.10 \ +0.22}$
& $14.60_{-0.26 \ -0.47}^{+0.38 \ +1.30}$ & $0.64_{-0.02 \ -0.05}^{+0.06 \ +0.21}$ \\

$~$ & $~$ & $~$ & $~$ & $~$ & $~$ \\
\hline
\end{tabular}}
\end{table}

The results we get by applying the likelihood analysis presented above are resumed in Table 1 where we also give the estimated values of other physically interesting quantities, namely the age of the universe $t_0$ and the transition redshift $z_T$. This latter quantity is defined by the condition $q(z_T) = 0$, being $q = -\ddot{a} a/\dot{a}^2$ the deceleration parameter, and, for the general case of the QSDM model, it is given by\,:

\begin{equation}
z_T = \left [ - \frac{(1 + 3 w)(1 - \Omega_b -
\Omega_{DM})}{\Omega_b + (1 - \varepsilon) \Omega_{DM}} \right
]^{- \frac{1}{3 w}} - 1 \ . \label{eq: zt}
\end{equation}
Since the uncertainties on the model parameters are not Gaussian distributed, a naive propagation of the errors is not possible. We thus estimate the 68$\%$ and 95$\%$ confidence ranges on the derived quantities by randomly generating 20000 points $(\Omega_{DM}, w, \varepsilon)$ using the marginalized likelihood functions of each parameter (if not held fixed) and then deriving the likelihood function of the derived quantity. Although not statistically well motivated, this procedure gives a conservative estimate of the uncertainties which is enough for our aims. Let us now briefly discuss the results for each model.

\begin{figure}[pt]
\centerline{\psfig{file=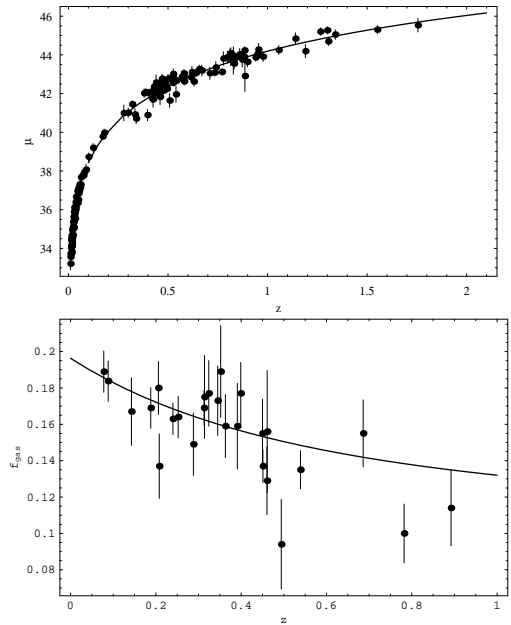,width=7cm}}
\vspace*{8pt}
\caption{Comparison among predicted and observed SNeIa Hubble diagram (upper panel) and $f_{gas}$ vs. $z$ relation (lower panel) for the best fit $\Lambda$CDM  model.}
\label{fig: lcdmfit}
\end{figure}

\subsubsection{$\Lambda$CDM}

Not surprisingly, the $\Lambda$CDM model gives an almost perfect fit to the dataset considered, as shown in Fig.\,\ref{fig: lcdmfit}. Having set from the beginning the Hubble constant, the only quantity to be determined is the dark matter density parameter $\Omega_{DM}$ that turns out to be severely constrained by the likelihood test. Adding to $\Omega_{DM}$ the baryon contribution $\Omega_b$ as set above, we get $\Omega_M = 0.319$ as best fit value, while the 95$\%$ confidence range is $(0.313, 0.335)$. It is worth comparing our result with those previously obtained by other authors. For instance, using only the SNeIa Gold dataset, Riess et al. \cite{Riess04} (hereafter R04) have found $\Omega_M = 0.29^{+0.05}_{-0.03}$. In the framework of the concordance model, a combined analysis of the CMBR anisotropy spectrum measured by WMAP, the power spectrum of SDSS galaxies, the SNeIa Gold dataset, the dependence of the bias on luminosity and the Ly$\alpha$ power spectrum lead Seljak et al. \cite{Sel04} (hereafter S04) to the estimate $\Omega_M = 0.284_{-0.060}^{+0.079}$ (at $99\%$ CL). Finally, fitting to the $f_{gas}$ data only with priors on both $h$ and $\Omega_b h^2$, but not imposing the flatness condition {\it ab initio}, A04 estimates $\Omega_M = 0.245^{+0.040}_{-0.037}$, while including the CMBR data, they get $\Omega_M = 0.26^{+0.06}_{-0.04}$. Overall, there is a very good agreement with our result. Nevertheless, it is worth noting that our best fit value is sistematically larger than that commonly quoted. This is partly due to having set $\Omega_b \simeq 0.05$ which is slightly larger than the fiducial value $\Omega_b \simeq 0.04$ often adopted. If we had set $h = 0.72$, the best fit value for $\Omega_M$ should be lowered by $\sim 0.01$ thus further reducing the difference with the {\it standard} result $\Omega_M \simeq 0.3$.

It is worth stressing that the substantial agreement among our estimated $\Omega_M$ and the previous results obtained using a variety of methods makes us confident that the likelihood analysis we have performed is correct and is not affected by some systematic errors. It is thus meaningful to apply this method to the other models presented in Sect.\,2.

Although not directly constrained by the fitting procedure, it is nonetheless interesting to compare the derived quantities reported in Table 1 with other estimates in literature. First, we consider the age of the universe $t_0$ whose best fit value turns out to be 13.27 Gyr. The $95\%$ confidence range is within the prior set on this quantity, but, as we will see later, this is not a general result. Most of the more recent estimates of $t_0$ have been obtained as a byproduct of fitting the $\Lambda$CDM model to a combination of different datasets and are thus rigorously model dependent. For instance, Tegmark et al. \cite{Teg03} (hereafter T04) give $t_0 =13.24_{+0.41}^{+0.89} \ {\rm Gyr}$, while Rebolo et al. \cite{VSA} find $t_0 = 14.4^{+1.4}_{-1.3}\ {\rm Gyr}$ (all at $1 \sigma$ level). Both these estimates agree with $t_0 = 13.6 {\pm} 01.19 \ {\rm Gyr}$ obtained by S04 that is the most comprehensive analysis. The accordance of our estimated $t_0$ with these results is not surprising due to the fact that we are using the same model and have yet obtained a similar value for $\Omega_M$.

\begin{figure}[pt]
\centerline{\psfig{file=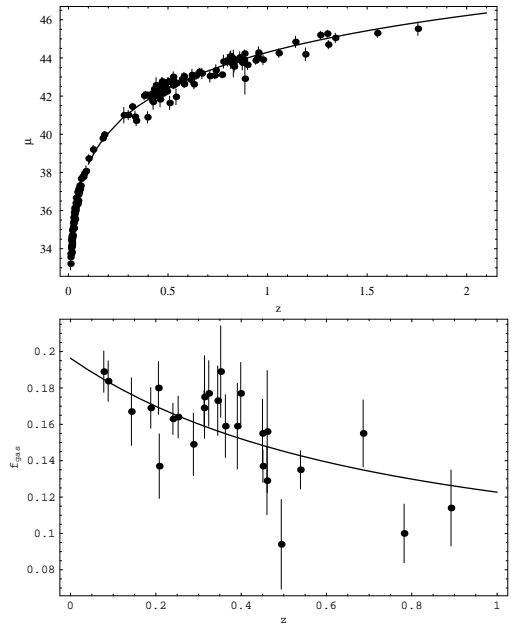,width=7cm}}
\vspace*{8pt}
\caption{The same as in Fig.\,\ref{fig: lcdmfit} but for the $\Lambda$SDM model.}
\label{fig: lsdmfit}
\end{figure}

It is therefore more interesting to consider the constraints on the transition redshift $z_T$. Since, for a flat $\Lambda$CDM model, $z_T$ only depends on $\Omega_M$, constraining $z_T$ is equivalent to constrain $\Omega_M$. Unfortunately, it is difficult to derive this quantity directly from the data even if some attempts have been made. Fitting the phenomenological parametrization $q(z) = q_0 + dq/dz|_{z = 0} z$ to the SNeIa Hubble diagram, R04 have found $z_T = 0.46 {\pm} 0.13$ (at $1 \sigma$ level) which is in marginal agreement with our $95\%$ confidence range. Since it is not clear what is the systematic error induced by the linear approximation of $q(z)$, which only works over a limited redshift range, we do not consider a serious shortcoming of the model the $1 \sigma$ disagreement between the R04 estimate of $z_T$ and the one reported in Table 1.

\subsection{$\Lambda$SDM}

Let us now consider the results obtained for the $\Lambda$SDM model in which the cosmic acceleration is still driven by the cosmological constant $\Lambda$, but the total dark matter content is made out by dust\,-\,like CDM particles and a fluid of strings. As clearly shown by Fig.\,\ref{fig: lsdmfit}, the model is able to fit very well both the SNeIa Hubble diagram and the $f_{gas}$ data. Moreover, the estimated $\Omega_{DM}$ (and thus $\Omega_M$) is in perfect agreement with that obtained for the $\Lambda$CDM model and hence with all other results discussed before. This could be qualitatively explained by noting that a fluid of strings is unable to drive cosmic acceleration even if it has negative pressure. Therefore, the amount of dark energy needed to accelerate the universe is the same as in the case of the $\Lambda$CDM model so that $\Omega_Q = 1 - \Omega_M$ (and hence $\Omega_M$) must be the same.

\begin{figure}[pt]
\centerline{\psfig{file=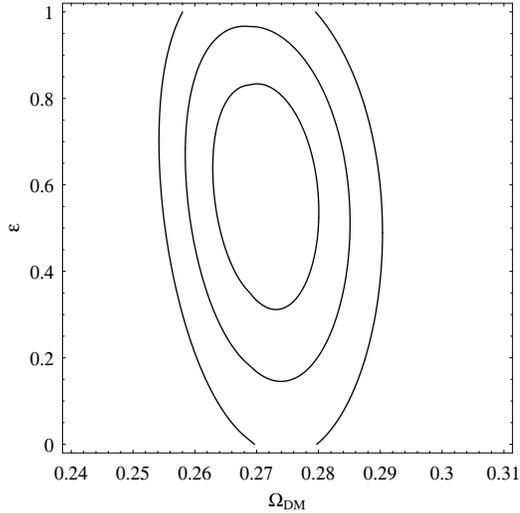,width=7cm}}
\vspace*{8pt}
\caption{$68\%$, $95\%$ and $99\%$ confidence contours in the $(\Omega_{DM}, \varepsilon)$ plane for the $\Lambda$SDM model.}
\label{fig: lsdmconts}
\end{figure}

The most striking result is, however, the constraint on $\varepsilon$, i.e. the fraction of dark matter represented by the fluid of strings. The best fit value turns out to be 0.59 and, what is more important, the value $\varepsilon = 0$ is safely excluded at more than $2 \sigma$ level. Looking at Fig.\,\ref{fig: lsdmconts}, where the likelihood contours in the plane $(\Omega_{DM}, \varepsilon)$ are plotted, shows that $\varepsilon$ and $\Omega_{DM}$ are positively correlated (even if weakly) so that it is not possible to reduce $\varepsilon$ without decreasing the total dark matter content. We may thus safely conclude that is possible to fit cosmological data with a significant fraction of the dark matter content in the form of a fluid of strings rather than CDM particles. This conclusion is reflected in the estimated $\Omega_b + \Omega_{CDM}$ that turns out to be much smaller than in the $\Lambda$CDM case, the best fit value being 0.154. This result may have interesting implications. Let us remember that a possible method to estimate $\Omega_M$ consists in estimating the mass\,-\,to\,-\,light ratio $M/L$ of clusters of galaxies and then integrating over the clusters luminosity function. Applying this method usually gives $\Omega_M \simeq 0.16$ (see, for instance, \cite{B00} and references therein) in striking disagreement with the results from tests probing cosmological scales (as SNeIa Hubble diagram and CMBR anisotropy spectrum). If we assume that the strings have a negligible mass\footnote{Note that this by no means imply that $\Omega_s$ is negligible since this is an {\it energy} rather than a {\it mass} density parameter. This could be best understood considering the case of radiation. Photons have zero rest mass, but nonetheless $\Omega_{rad}$ does not vanish today and was dominant in the past.} (which is a reasonable hypothesis \cite{Vil85}), we may qualitatively conclude that the method outlined above should give an estimate of $\Omega_b + \Omega_{CDM}$ rather than $\Omega_M$ since it is unable to weigth the contribution of the fluid of strings. Actually, the high value of $\varepsilon$ found has profound implications also at a galactic scales as it will be discussed in much detail later.

Although the picture of the today universe is similar to that of the $\Lambda$CDM model (i.e. the $\Lambda$ term dominates over the matter one), the presence a non\,-\,negligible fluid of strings alters the dynamics of the universe introducing, for instance, a fluid of strings dominated period. As a consequence, the age of the universe (which is sensitive to the full evolutionary history) is significantly changed with the best fit value $t_0 = 15.51 \ {\rm Gyr}$ that is outside the $95\%$ range obtained for the $\Lambda$CDM model. Moreover, this values is also larger than our best fit prior $t_0 = 13.9 \ {\rm Gyr}$ based on globular clusters and therefore model independent. However, $t_0$ depends linearly on $h^{-1}$ so that it is possible to decrease its value by simply increasing $h$ without changing the other model parameters (and still obtaining a very good fit to the data). For instance, using $h = 0.72$ gives $t_0 = 14.3 \ {\rm Gyr}$ in good agreement with the Rebolo et al. \cite{VSA} estimate quoted above. On the other hand, $t_0$ should be used to discriminate between the $\Lambda$CDM and $\Lambda$SDM model since they equally fit the same dataset, but predict significant different values for this quantity. Therefore, an accurate and model independent estimate of the age of the universe should make it possible to conclusively select one of the two models.

The presence of the fluid of strings also affects the transition redshift which turns out to be much higher (0.96 vs. 0.62 for the best fit values) than in the $\Lambda$CDM case (and hence more in disagreement with the estimate of R04). This result could be qualitatively explained considering that, in order to have deceleration, the universe must be dominated by dust matter. Therefore, introducing a non\,-\,negligible fluid of strings component, delays the onset of dust matter domination and thus increases $z_T$. If a reliable determination of this parameter were available in the future, we should obtain a further tool to confirm or reject the presence of the fluid of strings.

\subsubsection{QCDM}

\begin{figure}[pt]
\centerline{\psfig{file=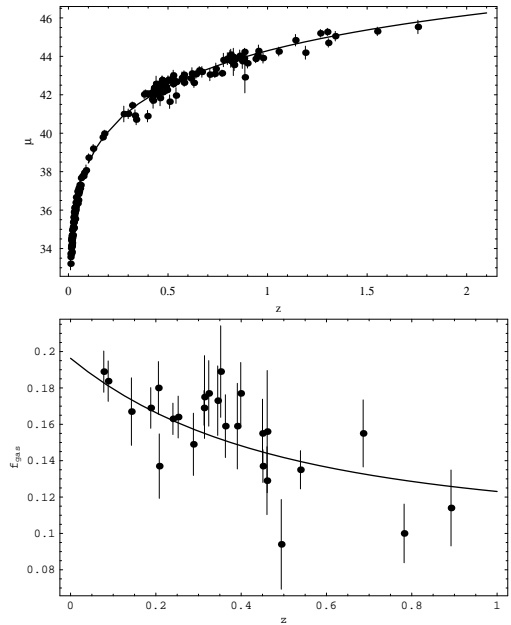,width=7cm}}
\vspace*{8pt}
\caption{The same as in Fig.\,\ref{fig: lcdmfit} but for the QCDM model.}
\label{fig: qcdmfit}
\end{figure}

Relaxing the hypothesis $w = -1$ but still keeping it constant and assuming again that there is no fluid of strings, we get the QCDM model where there are two parameters to be constrained, namely the dark matter density parameter $\Omega_{DM}$ and the barotropic factor $w$. The best fit is obtained for $(\Omega_{DM}, w) = (0.270, -1.28)$ and it is shown in Fig.\,\ref{fig: qcdmfit}, while we refer to Table 1 for the constraints on the parameters. In particular, we note that the results on $\Omega_{DM}$ (and hence on $\Omega_M$) are in perfect agreement with those obtained for the $\Lambda$CDM model so that we do not discuss anymore this parameter.

\begin{figure}[pt]
\centerline{\psfig{file=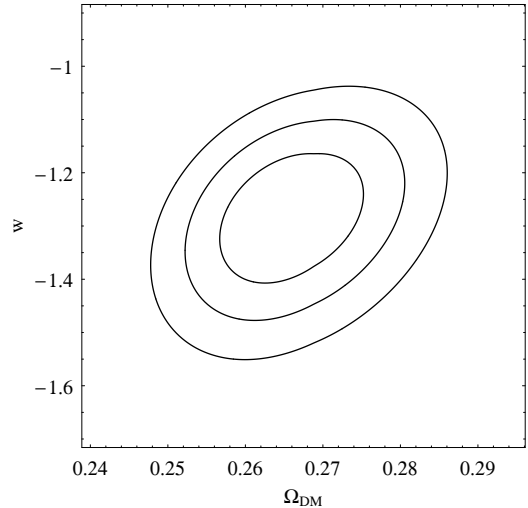,width=7cm}}
\vspace*{8pt}
\caption{The same as in Fig.\,\ref{fig: lsdmconts} but for the QCDM model.}
\label{fig: qcdmconts}
\end{figure}

It is more interesting to look at the constraints on the barotropic factor. First, we note that values of $w < -1$ are clearly preferred, while $w \ge -1$ is excluded at more than $95\%$ level. In particular, the cosmological constant is ruled out by the likelihood analysis. Moreover, Fig.\,\ref{fig: qcdmconts} shows that $w$ and $\Omega_{DM}$ are positively correlated so that increasing $w$ is only possible by unrealistically increasing the matter content. Surprising as it is, this result is however in agreement with previous analyses. Combining WMAP anisotropy spectrum with large scale structure clustering data and an old compilation of SNeIa, Spergel et al. \cite{WMAP} have found $w = -0.98 {\pm} 0.12$ when dropping the prior $w > -1$. Repeating the same analysis but using the Gold SNeIa sample, R04 have found $w = -1.02_{-0.19}^{+0.13}$. T04 added to the above dataset the power spectrum determined from the SDSS galaxy sample thus deriving $w = -0.72^{+0.34}_{-0..27}$. Finally, the comprehensive analysis of S04 gives $w = -1.080_{-0.198}^{+0.149}$ (all results given at $1 \sigma$). All these estimates agree among each other and our one, although we note that our best fit value is significantly smaller. Moreover, our result is the only one excluding the $\Lambda$CDM model at more than $99\%$ level considering the data which have been used.

Having pushed downward the confidence range for $w$ affects the predicted age of the universe. The maximum likelihood value turns out to be $t_0 = 14.50 \ {\rm Gyr}$, while, at $95\%$ level, $t_0$ lies between 14.24 and $14.79 \ {\rm Gyr}$. These values turns out to be higher than those for the $\Lambda$CDM model, but lower (even if in agreement at the $95\%$ level) than what is predicted in the $\Lambda$SDM case. As yet noted above, a comparison with the results obtained for the QCDM model by T03 and S04 suggests that including other kind of data (namely the CMBR anisotropy spectrum and the galaxy power spectrum) pushes upwards the constraints on $w$ with values closer to the $\Lambda$CDM ones. A similar conclusion also holds for the age of the universe. For instance, Tegmark et al. give $t_0 = 13.53_{-0.65}^{+0.52}$ which is, however, in agreement with our estimate. Note also that our value could be reconciled with the T03 estimate by increasing the value of $h$ from 0.664 to the value 0.71 used by these authors.

As a final remark, we note that the transition redshift $z_T$ is only weakly affected by relaxing the hypothesis $w = -1$ as it is witnessed by the good agreement between the estimates reported for the two cases. Indeed, $z_T$ mainly depends on the balance between $\Omega_M$ and $\Omega_Q$ so that, being $\Omega_{DM}$ and $\Omega_b$ the same in the two models, the resulting $z_T$ are naturally concordant.

\begin{figure}[pt]
\centerline{\psfig{file=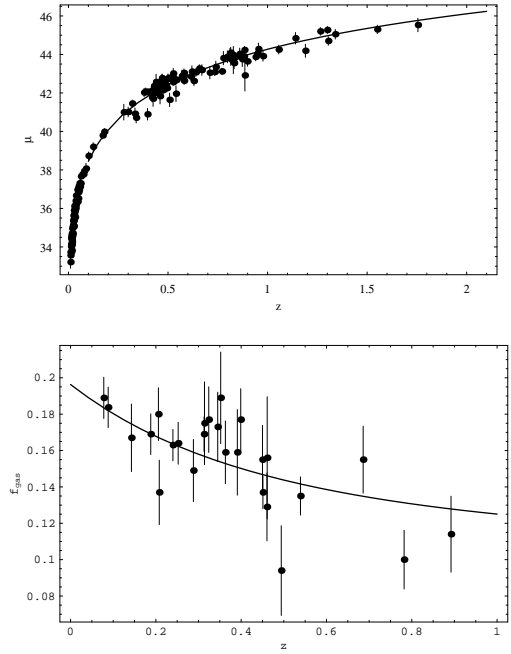,width=7cm}}
\vspace*{8pt}
\caption{Same as Fig.\,\ref{fig: lcdmfit} but for the QSDM model.}
\label{fig: qsdmfit}
\end{figure}

\subsubsection{QSDM}

Finally, we discuss the results for the general QSDM model where we relax the hypotheses on both the barotropic factor $w$ and the fraction $\varepsilon$ of dark matter made out of fluid of strings. Having now three parameters to constrain ($\Omega_{DM}, w, \varepsilon)$, it is not
surprising that the confidence ranges enlarge. While this is only a minor effect for what concerns $\Omega_{DM}$, there is a significant weakening of the constraints on $w$ and a dramatic impact on $\varepsilon$ on which we are only able to give upper limits. Nevertheless, the results are quite interesting. In particular, the best fit is obtained for a model that is very similar to the best fit QCDM one with the same content of dark matter (i.e. the same $\Omega_{DM}$), almost the same $w$ (-1.21 vs. -1.28) and no fluid of strings. Not surprisingly, the fitting to the data is perfect as shown in Fig.\,\ref{fig: qsdmfit}.

\begin{figure}[pt]
\centerline{\psfig{file=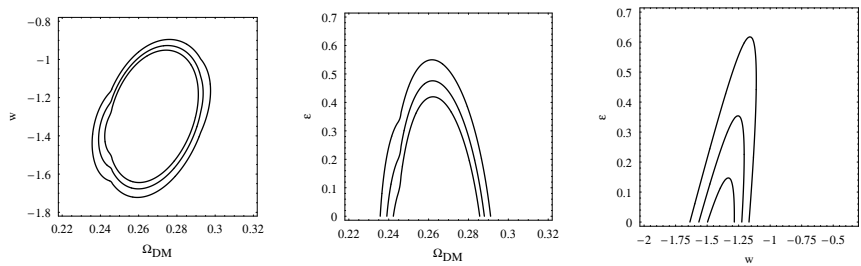,width=12cm}}
\vspace*{8pt}
\caption{The same as in Fig.\,\ref{fig: lsdmconts} but for the QSDM model. In each panel, the value of the parameter not shown on the axes is set to its best fit value reported in Table 1.}
\label{fig: qsdmconts}
\end{figure}

Although quite similar to the QCDM model for the best fit values, the QSDM case is however different for what concerns the constraints on its parameters because of the presence of a third quantity to be determined. This is pictorially shown in Fig.\,\ref{fig: qsdmconts} where we plot the projections of the likelihood function on the three orthogonal planes in the parameter space. It is worth noting that $\Omega_{DM}$ is the parameter least affected by the addition of a third quantity so that the resulting constraints are only marginally weakened and the $68\%$ and $95\%$ confidence ranges nicely overlap with those obtained for all the models considered above. As an obvious consequence, our estimate for $\Omega_{M}$ (that is, $\Omega_b + \Omega_{DM}$) is also in agreeement with the other results quoted above. Much care is needed when considering the constraints on the derived $\Omega_b + \Omega_{CDM}$. Since the best fit model has a vanishing $\varepsilon$, the maximum likelihood value turns out to be almost the same as those for the $\Lambda$CDM and QCDM model that also have no dark matter in fluid of strings. Nevertheless, the upper limits on $\varepsilon$ leads to strongly asymmetric constraints on the standard matter density parameter $\Omega_b + \Omega_{CDM}$ so that, at the $95\%$ level, also values as low as 0.166 are possible thus being more in line with the prediction of the $\Lambda$SDM model.

As a general rule, the QSDM model is indeed very similar to the QCDM one, having almost the same matter content (and hence the same dark energy content) and constraints on the barotropic factor $w$ which agree very well with those of the QCDM model, but are less stringent. As a result, we get also concordant estimates of both the age of the universe and the transition redshift so that we do not comment again on these quantities.

\subsection{Comparing the models}

The results of the likelihood analysis discussed above have shown that all the models we have considered are able to fit well the same dataset. The next natural step is wondering which is the better one. To answer this question is not an easy task. Combining different datasets requires the introduction of the pseudo\,-\,$\chi^2$ defined in Eq.(\ref{eq: defchi}) that is not the same as the $\chi^2$ commonly used in statistical analysis. Moreover, the models have a different number of parameters to be constrained. As such, it makes no sense comparing the models on the basis of the $\chi^2$ best fit value. To overcome this difficulty, Liddle \cite{Lid04} have proposed to resort to the {\it information criteria} that are widely used in other branches of science (such as medical pathologies), but poorly known in astrophysics. In particular, Liddle have proposed to use the {\it Akaike information criterion} (AIC) defined as \cite{Ak74}\,:

\begin{equation}
{\cal{A}} \equiv -2 \ln{{\cal{L}}} + 2 k
\label{eq: aic}
\end{equation}
and {\it Bayesian information criterion} (BIC) \cite{S78}\,:

\begin{equation}
{\cal{B}} \equiv -2 \ln{{\cal{L}}} + k \ln{N}
\label{eq: bic}
\end{equation}
with ${\cal{L}}$ the likelihood evaluated for the best fit parameters, $k$ the number of model parameters and $N$ the total number of points in the dataset used. Since both AIC and BIC explicitly takes into account the number of parameters, it is meaningful to compare the models on the basis of the values of these quantities. The lower is ${\cal{A}}$ or ${\cal{B}}$, the better the model is. Unfortunately, as discussed by Liddle, it is not an easy task to decide what is the better information criterion to be adopted so that we conservatively report in Table 2 the values of both ${\cal{A}}$ and ${\cal{B}}$ for the best fit model parameters. The models are quite similar in their AIC and BIC values, as it is expected since they equally fit the same dataset. Nevertheless, it is possible to rank them according to ${\cal{A}}$ or ${\cal{B}}$. Independently on what criterion is used, the QCDM model turns out to be the preferred one followed by the QSDM or the $\Lambda$SDM depending on which information criterion (AIC or BIC respectively) is adopted. This simple comparison leads to two quite interesting conclusions.

\begin{table}[pt]
\tbl{Number of parameters $k$, AIC ${\cal{A}}$ and BIC ${\cal{B}}$ for the models discussed in the text.}
{\begin{tabular}{|c|c|c|c|}
\hline
Id & $k$ & ${\cal{A}}$ & ${\cal{B}}$ \\
\hline \hline
$\Lambda$CDM & 1 & 225.687 & 228.896 \\
$\Lambda$SDM & 2 & 215.830 & 222.249 \\
QCDM & 2 & 210.933 & 217.352 \\
QSDM & 3 & 213.456 & 223.084 \\
\hline
\end{tabular}}
\end{table}

First, there is a clear evidence favouring models others than the concordance $\Lambda$CDM. In particular, the AIC suggests that phantom like models have to be preferred so that a violation of the strong energy condition is unavoidable to explain the cosmic acceleration in the framework of constant $w$ dark energy models.

A second and perhaps more interesting result is that the $\Lambda$SDM model is preferred over the $\Lambda$CDM model and only marginally disfavoured with respect to the QCDM one. Therefore, introducing a fluid of strings as a component of the dark matter term makes it possible not only to better fit the data, but it could also be a viable alternative to phantom like models. Surprising as it is, this result is quite encouraging and motivates further study.

\section{The weak field limit}

The results discussed above has shown that introducing a fluid of strings in the dark matter budget does modify the dynamics of the universe, but gives rise to a model that is still in agreement with the SNeIa Hubble diagram and the data on the gas mass fraction in galaxy clusters and predict an age of the universe which is not unreasonable. Moreover, both the Akaike and Bayesian information criteria quantitatively indicate that the $\Lambda$SDM model has to be preferred over the concordance $\Lambda$CDM.

It is worth stressing that, for the $\Lambda$SDM model, the fraction of dark matter represented by the fluid of strings is nearly dominant so that the density parameter of the standard (baryons + CDM) is significantly smaller with a best fit value as low as 0.154. It is easy to understand that lowering $\Omega_{CDM}$ have a profound impact at galactic scales. Indeed, since we have assumed that the strings have a very small (if not vanishing) mass, a small $\Omega_{CDM}$ automatically implies less massive dark matter haloes. It is thus worth wondering whether such light haloes may still fit the rotation curves of spiral galaxies. Naively, one should think that the answer is negative since values of $v_c(r)$ larger than those predicted on the basis of luminous matter only naturally invoke massive haloes. But this is only true in a Newtonian gravitational potential. This is no more the case in a $\Lambda$SDM model. Considering the weak field limit, Soleng has shown that the gravitational potential for a pointlike mass $m$ embedded into a halo of fluid of strings is given by\footnote{It is importnat to stress that we are considering finite length strings rather than a network of cosmic strings with infinite length. Indeed, in this second case, it is still not clear how to compute the gravitational potential in the weak field limit.} \cite{soleng}\,:

\begin{equation}
\Phi(r) = - \frac{c^2}{2} \left [ \frac{\xi_1}{r} + \frac{\alpha}{\alpha - 2} \left ( \frac{\xi_2}{r} \right )^{2/\alpha} \right ]
\label{eq: phisol}
\end{equation}
where $c$ is the speed of light, $\xi_{1,2}$ are two integration constants and $\alpha = -\rho/p$. Since, for our model, it is $\alpha = 3$, imposing the condition that $\Phi(r)$ reduces to the usual Newtonian potential in the case $r << \xi_1, \xi_2$, we get $\xi_1 = 2Gm/c^2$ so that, with simple algebra, we can rewrite Eq.(\ref{eq: phisol}) as\,:

\begin{equation}
\Phi(r) = - \frac{G m}{r} \left [ 1 + \left ( \frac{r}{\xi} \right )^{1/3} \right ]
\label{eq: phipoint}
\end{equation}
where we have defined a new scalelength $\xi$ including all the constants. Eq.(\ref{eq: phipoint}) gives the potential for a pointlike mass. In order to generalize this result to the case of an extended system, we may divide the system in infinitesimal mass elements $dm$ and sum up the contributions to get the total potential. Assuming for the sake of simplicity spherical symmetry, the gravitational potential of an extended halo is thus\,:

\begin{equation}
\Phi(r) = - \frac{G M(r)}{r} \left [ 1 + \left ( \frac{r}{\xi} \right )^{1/3} \right ]
- 4 \pi G \int_{0}^{r}{\rho(r') r' \left [ 1 + \left ( \frac{r'}{\xi} \right )^{1/3} \right ] dr'}
\label{eq: phiext}
\end{equation}
where $M(r)$ and $\rho(r)$ are respectively the halo mass and the density profile. The first term in Eq.(\ref{eq: phiext}) represents the contribution to the gravitational potential of the mass within the radius $r$,  while the second one takes into account the mass outside this radius. The circular velocity depends on the force acting on the star orbiting at distance $r$ from the halo centre and is thus only determined by the first term in Eq.(\ref{eq: phiext}) as a result of the Gauss theorem \cite{BT87}. A straightforward generalization of the standard formula then gives in this case\,:

\begin{equation}
v_c^2(r) = \frac{G M(r)}{r} \left[ 1 + \frac{2}{3} \left ( \frac{r}{\xi} \right )^{1/3} \right ] \ .
\label{eq: vcstrings}
\end{equation}
Comparing this result with the Newtonian formula $v_c^2(r) = G M(r)/r$ shows that the circular velocity is higher because of the additive term $(r/\xi)^{1/3}$. Qualitatively, Eq.(\ref{eq: vcstrings}) shows that the fluid of strings effectively works as a fictitious source distributed with a mass profile $2M(r)/3 \times (r/\xi)^{1/3}$. Because of this additional effective source, it is possible to get the same value of $v_c(r)$ as in the Newtonian case with a smaller value of the CDM halo mass. Summarizing, introducing the fluid of strings modifies the gravitational potential in such a way that less CDM is necessary to get a given value of $v_c(r)$.

The reduction of the quantity of CDM needed to fill the dark matter haloes is consistent with what is expected from our previous estimate of $\Omega_b + \Omega_{CDM}$ in the $\Lambda$SDM model. Indicating with $\Omega_{CDM}^s$ ($\Omega_{CDM}^N$) the CDM density parameter for the $\Lambda$SDM ($\Lambda$CDM) model, we may qualitatively write\,:

\begin{displaymath}
\frac{\Omega_{CDM}^s}{\Omega_{CDM}^N} \propto \frac{M_{CDM}^s}{M_{CDM}^N}
\end{displaymath}
where $M_{CDM}^s$ ($M_{CDM}^N$) is the typical mass in CDM particles in a $\Lambda$SDM ($\Lambda$CDM) model. Naively speaking, in order to get the same value of the rotation curve at, for instance, the virial radius\footnote{The virial radius is defined such that the mean mass density within $r_{vir}$ is $\delta_{th}$ times the mean matter density $\bar{\rho} = \Omega_M \rho_{crit}$, with $\delta_{th}$ the critical overdensity for the gravitational collapse of density perturbations.} $r_{vir}$, the halo mass in the $\Lambda$SDM case must be smaller than in the $\Lambda$CDM case by an amount that is of the order of magnitude of the correction term in Eq.(\ref{eq: vcstrings}) evaluated at $r_{vir}$ so that it is\,:

\begin{displaymath}
\frac{\Omega_{CDM}^s}{\Omega_{CDM}^N} \sim \frac{2}{3} \left ( \frac{r_{vir}}{\xi} \right )^{1/3} .
\end{displaymath}
According to the results in Table 1, $\Omega_{CDM}^s/\Omega_{CDM}^N \simeq 1/2$ so that, from the above relation, we get $\xi \sim (4/3)^3 r_{vir}$. Such high values of $\xi$ also ensures that the gravitational potential is practically the same as the classical Newtonian one in the galactic regions dominated by the visible components where the rotation curve is well fitted by using the standard formulae. It is worth stressing, however, that this encouraging result needs to be further investigated by a careful fitting to the rotation curves of observed galaxies. This is outside the aim of this paper, but will be presented in a forthcoming work.

\section{Conclusions}

Shedding light on the dark side of the universe is a very difficult, but also very attractive challenge of modern cosmology. The nature and the fundamental properties of the two main ingredients of the cosmic pie, namely the dark energy and the dark matter, are still substantially unknown and it is, indeed, this wide ignorance that justifies and motivates the impressive amount of theoretical models proposed to explain the observed astrophysical evidences. Moving in this framework, we have considered the dark matter as made out not only of massive dustlike CDM particles, but also of a fluid of strings of cosmic origin with an equation of state $w_s = -1/3$. Starting from this idea, we have considered four cosmological models comprising four components, namely dustlike baryons and CDM, fluid of strings and dark energy with constant barotropic factor $w$. Two of these four models ($\Lambda$SDM and QSDM) have a non vanishing fraction of dark matter in the form of a fluid of strings, while in two models ($\Lambda$CDM and $\Lambda$SDM) the energy budget is dominated by the $\Lambda$ term. Our main results are briefly outlined as follows.

\begin{enumerate}

\item{All the models are able to fit the data on the SNeIa Hubble diagram and the gas mass fraction in galaxy clusters with very good accuracy. In particular, it is remarkable that the total dark matter density parameter $\Omega_{DM} = \Omega_{CDM} + \Omega_s$ is very well constrained and turns out to be the same in all models. When the assumption $w = -1$ is relaxed, the dark energy barotropic factor is constrained to be in the region $w < -1$ so that phantom like models are clearly preferred with a disturbing violation of the strong energy condition. It is worth noting that present data do not require phantom dark energy since they can be equally well fit by models with the cosmological constant $\Lambda$ driving the accelerated expansion. Discriminating among the different possibilities will need a large sample of high redshift SNeIa such as those that should be available with the SNAP satellite mission \cite{SNAP}.}

\item{According to both the AIC and BIC, the QCDM model is statistically preferred over the other considered possibilities and this is not an unexpected result. However, this is obtained to the price of admitting phantom dark energy which is affected by serious theoretical difficulties. On the other hand, the $\Lambda$SDM model is preferred over the popular concordance $\Lambda$CDM scenario and is only slightly disfavoured with respect to the QCDM one. The good fit to the data and the graceful feature of avoiding to enter the realm of ghosts makes this model a good compromise between observations and theory and we therefore consider it as our final best choice.}

\item{The $\Lambda$SDM model predicts that a significant fraction ($\varepsilon \simeq 59\%$) of the dark matter is made out by a fluid of strings so that the standard matter density parameter $\Omega_b + \Omega_{CDM}$ is only half of the fiducial value in the concordance scenario (0.15 vs 0.30). Since we may assume that the fluid of strings is massless (or nearly so), we should expect a corresponding decrease of the mass of galactic dark haloes. If the gravitational potential is still Newtonian, decreasing the CDM halo mass should lead to lower values of the circular velocity in the outer dark matter dominated regions of galaxies. This is not the case since, in the weak field limit, the $\Lambda$SDM model gives rise to a modification of the gravitational potential. As a result, the circular velocity due to a mass $M(r)$ is higher than in the classical case so that less massive haloes are necessary to give the observed values of $v_c(r)$. Moreover, a very qualitative calculation suggests that the typical value of the scalelength over which deviations from Newtonian formulae cannot be neglected is sufficiently high that the inner luminous matter dominated rotation curve is unaltered.}

\end{enumerate}

These encouraging results motivate further studies of the $\Lambda$SDM model. To this end, there are two different routes connected to two different features of the model which can be followed.

First, because of its scaling with the redshift as $\rho_s \propto (1 + z)^2$, that is intermediate between that of CDM and that of $\Lambda$, a new era dominated by a fluid of strings is predicted in the expansion history of the universe. It is thus worth investigating how this imprints on the CMBR anisotropies in order to see whether the spectrum measured by WMAP is still accurately reproduced. To this regard, it is worth noting that the attempts recently made to constrain the cosmic strings contribution to the CMBR spectrum \cite{Fraisse} may not be extended to our case since they refer to a network of cosmic strings rather than clouds of finite length strings. Less theoretically demanding, but more observationally ambitious is the possibility to test the proposed scenario on the basis of the transition redshift $z_T$. As Table 1 shows, for the $\Lambda$SDM model, $z_T$ is significantly higher than in the other cases so that a model independent estimate of this quantity could be a powerful discriminating tool.

One of the most peculiar features of the $\Lambda$SDM model is the modified gravitational potential in Eq.(\ref{eq: phiext}) leading to the corrected circular velocity in Eq.(\ref{eq: vcstrings}). Having been obtained in the weak field limit, such correction should be tested at the scale of galaxies and clusters of galaxies thus offering the possibility to test the model at a very different level. To this aim, one should try fitting the rotation curve of spiral galaxies to see whether the problem of their flatness could be solved in this framework. Moreover, it is interesting to check how much the halo mass is reduced and to compare the reduction with respect to the classical Newtonian estimates with the decreasing of $\Omega_{CDM}$ obtained above. To this aim, low surface brightness (LSB) galaxies are ideal candidates since they are likely dark matter dominated so that systematic uncertainties on the luminous matter modelling have only a minor impact on the fitting procedure. Moreover, the stellar mass\,-\,to\,-\,light ratio of LSB galaxies is well constrained so that we may fix this quantity thus decreasing the degeneracy among the other parameters. Useful samples of LSB galaxies with detailed measurements of the rotation curve are yet available (see, for instance, \cite{dBB02}) so that this kind of test may be easily implemented. In this same framework, it is also interesting to consider the velocity dispersion curves in elliptical galaxies where recent studies seem to indicate a dark matter deficit \cite{Cap}.

Changing the gravitational potential does not only alter galaxies rotation curve, but also affects the clustering properties and thus leads to a different matter power spectrum. It is thus interesting to compare the predicted power spectrum with those measured from the SDSS galaxies in order to check the validity of the $\Lambda$SDM model. A similar comparison has been recently performed by Shirata et al. \cite{SSYS04} for two phenomenological modifications of the law of gravity. We stress, however, that their approach is purely empirical and, furthermore, assumes that the universe can still be described at large scales with the $\Lambda$CDM model. Since in order to compute the power spectrum, one also needs the background Hubble parameter, it is important to use an expression for $H(z)$ that is consistent with the proposed modification of gravity. For the $\Lambda$SDM model considered here, all the ingredients are at disposal so that a coherent calculation can be performed. It is worth noting that such a test is the only one capable of probing the model both at the galactic (through the gravitational potential) and cosmological (because of the use of $H$) scales at the same time.

As a concluding general remark, we would like to stress the need for tackling the dark matter and dark energy problem together taking care of what is the effect at the galaxy scale of any modification of the fundamental properties of one of these two components. In our opionion, this could be a valid approach in elucidating the problems connected to the dark side of the universe.

\section*{Acknowledgments}

We warmly thank R.W. Schmidt for having kindly furnished us in electronic form the data on the gas mass fraction in advance of publication.

\end{document}